# V-Shape Liquid Crystal-Based Retromodulator Air to Ground Optical Communications

**M. A. Geday, G. del Campo, A. Carrasco, N. Bennis, X. Quintana, F. J. López-Hernandez, and J. M. Otón**
Departamento de Tecnología Fotónica, ETSI Telecomunicación,
Universidad Politécnica de Madrid, Ciudad Universitaria,
Madrid, Spain

*This paper describes the use of a 2D liquid crystal retro-modulator as a free space,* wireless*, optical link. The retro-modulator is made up of a retro-reflecting corner-cube onto which 2 cascaded V-shape smectics liquid crystal modulators are mounted. The communication link differs with respect to more conventional optical links in not using amplitude (nor frequency) modulation, but instead state-of-polarisation (SOP) modulation known as Polarisation Shift Keying (PolSK). PolSK has the advantage over amplitude modulation, that it is less sensitive to changes in the visibility of the atmosphere, and increases inherently the bandwidth of the link. The implementation of PolSK both in liquid crystal based and in retro-modulated communication are novelties.*

## INTRODUCTION

Conventional wireless optical communications feature a number of advantages over radio frequency communications, hereunder low divergence of the communication channel and beam steering which prevents eavesdropping, allows for smaller receivers, and allows for an increased bandwidth for deep space communications [1]. However in near earth orbit satellites and high altitude balloons, where weight and power consumption are limiting factors, the implementation

The authors should like to thank Ingenieria y Servicios Aeroespaciales (INSA) for financial support and Prof. Dobrowski, Military University of Technology, Warsaw, Poland for providing the V-shape smectics.

Address correspondence to M. A. Geday, Department Tecnologia Fotonica, ETSI Telecomunicacion, Universidad Politecnica de Madrid, Ciudad Universitaria s/n, Madrid, E 28040, Spain. E-mail: morten@tfo.upm.es

including laser source, power supply and tracking system, compromises the airborne weight and becomes a problem. With this in mind we have designed a light weight low power retro-modulator based on V-shape liquid crystals (Fig. 1).

The airborne system consists of a retro-reflecting prism onto which a liquid crystal modulator is mounted (Fig. 2). When light enters a retro-reflecting prism it will be reflected three times on the internal surfaces of the prism and will return in the direction it originated from. Hence, by using a retro-reflecting mirror, any tracking device mounted on the airborne unit becomes superfluous.

The idea of using liquid crystals and retro-reflectors is by no means new [2], it has been studied by numerous research groups in the past. Most significantly, Swenson *et al.* [3] reported a data transfer speed of 20 kbits, from an air balloon at an altitude at 32 km using a simple ON-OFF data transfer scheme with ferroelectric liquid crystals as the modulating element. They achieved their results by combining nine 1-inch retro-modulators (mounted in order to enlarge the field of view), using a 5 watts laser source (810 nm) and a 1.5 m telescope. More recently the general choice of technology for retro-modulation has become the multiple quantum well (MQW) modulators, which have a switching speed which makes GB communication possible, e.g., [4], and liquid crystals have been employed in the beam steering of the interrogating light beam [5]. It should however be noted that the active area of a MQW is less than 2 centimetres in diameter, and as the signal power in retro-modulation is proportional to the fourth power of the modulator diameter [6], the role of liquid crystal based retro modulators should not be completely disregarded, particularly in low visibility.

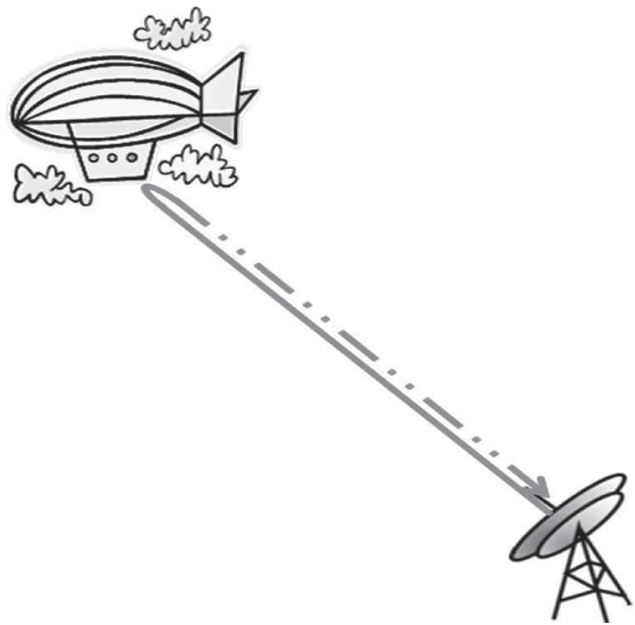

**FIGURE 1** The principle of retro-modulation: A laser beam is directed towards a cats-eye or retro-mirror, which is combined with an electro optical modulator. The retro-mirror returns the modulated beam towards the emitter, and thus an airborne laser is avoided.

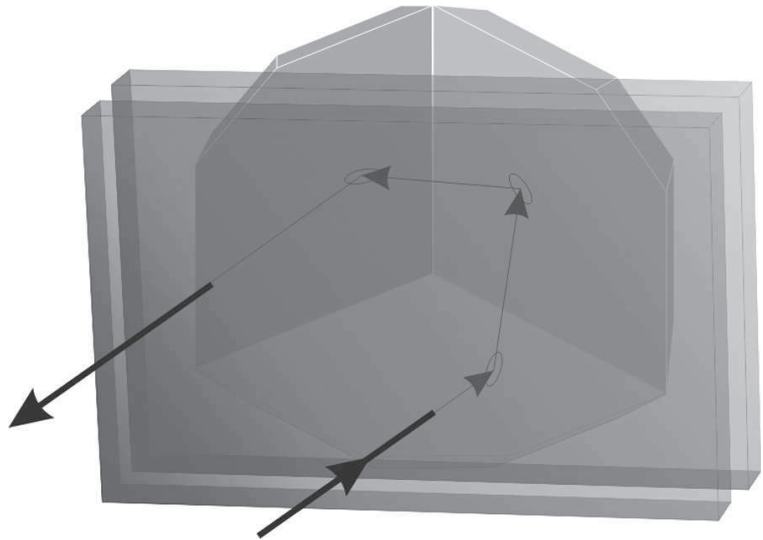

**FIGURE 2** A corner cube prism with two liquid crystal modulators mounted on top. The light passes through the liquid crystal once, bounces on the three internal surfaces of the retro-prism, and passes through the liquid crystal modulators to return in the direction of origin.

Wireless optical communication based using PolSK, is a well described technology [7]. In the simplest implementation the transmitter consists of a polarised laser input followed by two phase modulators. The two polarisations are then recombined using to create any SOP.

## IMPLEMENTATION

In our setup the ground system includes a laser, a tracking unit, and a receiver. The airborne liquid crystals modulate the upcoming light beam, and the retro-reflector prism reflects the light back towards the ground station where an in-house-built real time polarimeter determines the state of polarisation (SOP). One advantage of using SOP variation instead of amplitude variation, i.e., making the equivalent of a reflexive display, is that the detector will always receive a signal, albeit of varying polarisation. Furthermore, the bit rate can be increased by encoding several bits per symbol in the same light pulse, and since the signal will be coded as relative amplitude and phase differences, the system will be more robust towards temporal variations in visibility of the atmosphere.

Various configurations have been considered. The simplest, the amplitude modulation, where a single liquid crystal modulator is mounted between crossed polarisers is a perfect analogue to a reflective liquid crystal display. We disregarded this configuration *a priori* because this configuration has already been thoroughly studied by others [3], and the likelihood of increasing the bit rate significantly, simply by using a single analogue switching liquid crystal modulator with either nematic or V-shape smectic materials in order to generate various grey-level didn't seem scientifically interesting. Furthermore, the 4 passes through a dichroic sheet polariser would lead to an attenuation of signal that could become critical when working with retro-modulators in adverse atmospheric conditions.

Instead PolSK modulation was decided upon. Two configurations were considered. The first involved mounting a polarising sheet between the retro-reflector prism and the liquid crystal modulator, and then detect the state of polarisation on ground. The other was to transmit light with a know polarisation from ground, and then modulate it both before entering and after exiting the retro-reflector prism.

In the end the latter configuration was decided upon. The decision was based on two arguments. First of all powerful polarised laser sources are readily available, which means that the – even in the ideal case – 50% loss of intensity caused by an airborne polarising element, implicit in the first configuration, unattractive. Secondly passing through the modulator twice opened up the possibility of using either less birefringent, and as a rule less viscous, liquid crystals, or to use thinner, and hence faster switching liquid crystal cells. With the subsequently described choice of using V-shape smectics materials, the selection was made even easier as these materials require thin cells in order to be properly surface stabilised.

The next choice was to decide upon configuration of the liquid crystal material. Having decided upon PolSK, or more accurately multilevel PolSK, the option was between nematics and V-shape smectics liquid crystals; considering the response speed, V-shape smectics was the obvious choice. After testing the materials available in the laboratory it was decide to use the experimental mixture W212, synthesised at Military University of Technology, Warsaw, Poland. The cone-angle (close to 85°) and the birefringence ($\Delta n = 0.14$ at 632 nm) of the material made it possible to generate almost any light SOP, when passing through two modulators twice, as will be discussed below. The material was characterised electro-optically, and the switching behavior was determined (Fig. 3).

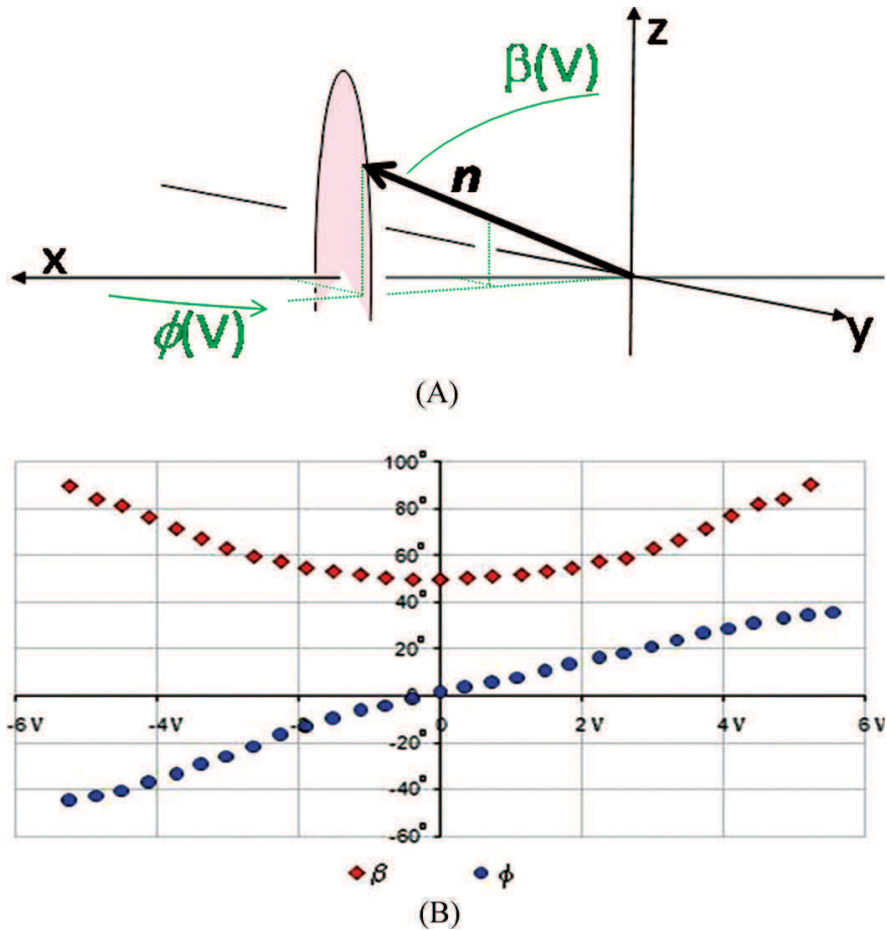

**FIGURE 3** (A) The electro-optical response of the employed V-shape smectic (W212) in terms of in-plane ($\phi$, with respect to smectic layer normal), and the out-of-plane ($\beta$, with respect to cell normal) movement of the director (**n**). The switching cone has been indicated in pink. (B) The values of $\beta$ and $\phi$ (measured with an arbitrary offset), where deduced from measurements of birefringence and azimuthal angle using a rotating linear analyser and circular polarised light [9].

The final choice concerned the original (upwards) polarisation of the interrogating beam. The initial decision fell upon circularly polarised light in order to make the light impinging on the retromodulator independent of the azimuthal (horizontal) orientation of the vehicle.

This preliminary study employs a 632 nm laser source. In a final application IR wavelengths (850 nm) shall be considered.

## MATHEMATICAL DESCRIPTION

In order to use the retromodulator for optical communications, it is necessary to deduce the switching voltages applied to the modulating liquid crystal cells, indifferently to the altitude and rotation of the retro-modulator. The altitude of the modulator affects the angle of incidence of the light, and the rotation affects the relative in-plane switching of the liquid crystals.

Assuming that the birefringence of the liquid crystal cells is uniaxial, the description of the light path is trivial, but cumbersome. The geometry is represented in Figure 4. It is complicated by fact that the balloon may be at any height and distance, i.e., at any *zenithal angle* $\gamma_o$, with respect to vertical ($z$), and at any rotation, *azimuthal angle* $\phi_0$, with respect to the horizontal pointing direction. Furthermore, the V-shape smectic liquid crystal switching has both an in plane ($\phi$) and an out of plane ($\beta$) component. Combine the four angles one can determine the angle between the molecular director, and the incoming light ($\Theta$), and hence the magnitude of the plano-birefringence ($\Delta n_{eff}$) [8] and the introduced phase shift ($\delta$). The inclination of the elliptic cross-section ($\alpha$), can be determined as the projection of the director onto the plane perpendicular to the interrogation light beam (x″, y′). Upon characterisation of the electro-optical response of the liquid crystal, both $\delta$ and $\alpha$, can be cast as functions of the three parameters $\gamma_o$, $\phi_0$ and the applied field ($V$).

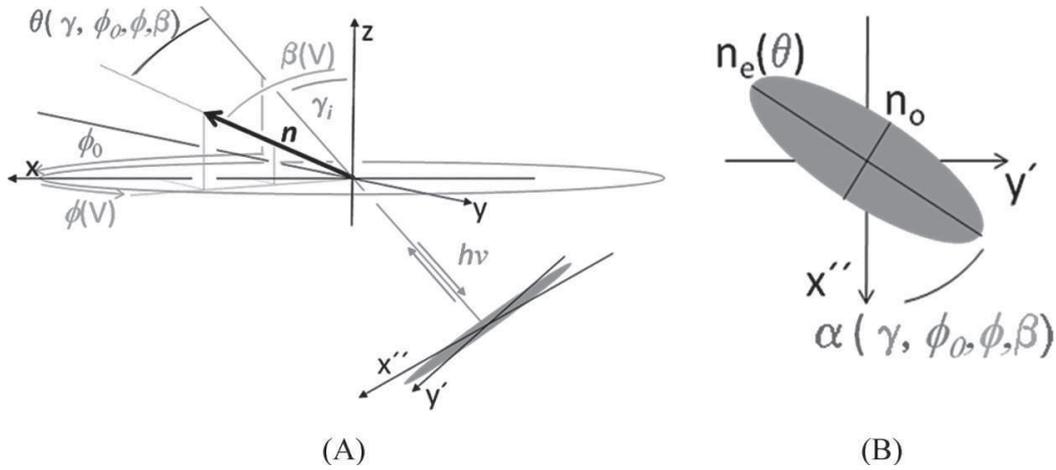

(A)             (B)

**FIGURE 4** (A) A geometrical description of the angles involved (red: the light impinging on the retromodulator, at an angle of incidence ($\gamma_i$); blue: the horizontal rotation of the retromodulator, with respect to pointing direction ($\phi_0$); green: the switching of the LC (in-plane $\phi(V)$ and out of plane $\beta(V)$). Combining $\gamma_i$ ; $\phi_0$, $\phi(V)$ and $\beta(V)$, one can deduce the angle between the incoming light and the LC director ($\Theta$) and thus the elliptic cross-section of the indicatrix as indicated in purple. All angles and coordinates are defined in the text. It should be noted that the $\gamma_i$, indicated in the figure is the *internal* angle of incidence, not the *external* angle of incidence, the azimuthal angle. The angles are shown for one LC modulator only. Combining two LCs mean that one of these will have an additional offset angle $\Delta\phi$ that must be added to the $\phi_0$ term. (B) The elliptic cross-section in the plane of SOP detector, and hence perpendicular to the interrogating beam, $\alpha$ is the angle between vertical projection of the pointing direction, and the major axis of the elliptic cross-section.

Knowing $\delta$ and $\alpha$ for the two liquid crystals the full light path can be described using Jones formalism:

$$\mathbf{J}_{out} = \mathbf{F}_\gamma \cdot \mathbf{R}_{-\alpha_1} \cdot \mathbf{M}_{\delta_1} \cdot \mathbf{R}_{\alpha_1} \cdot \mathbf{R}_{-\alpha_1} \cdot \mathbf{M}_{\delta_2} \cdot \mathbf{R}_{\alpha_2} \cdot$$
$$\mathbf{Mirr} \cdot \mathbf{R}_{\alpha_2} \cdot \mathbf{M}_{\delta_2} \cdot \mathbf{R}_{-\alpha_2} \cdot \mathbf{R}_{\alpha_1} \cdot \mathbf{M}_{\delta_1} \cdot \mathbf{R}_{-\alpha_1} \cdot \mathbf{F}_\gamma \cdot \mathbf{J}_{in}$$

Where $\mathbf{J}_{in}$ is the incoming light polarisation and the various matrices are defined as:

$$\mathbf{R}_\alpha = \begin{bmatrix} \cos(\alpha) & -\sin(\alpha) \\ \sin(\alpha) & \cos(\alpha) \end{bmatrix}, \quad \mathbf{M}_\delta = \begin{bmatrix} 1 & 0 \\ 0 & e^{-i\delta} \end{bmatrix},$$

$$\mathbf{F}_\gamma = \begin{bmatrix} 1 & 0 \\ 0 & \cos(\gamma_o - \gamma_i) \end{bmatrix}, \quad \mathbf{Mirr} = \begin{bmatrix} 1 & 0 \\ 0 & -1 \end{bmatrix}$$

$\mathbf{R}_\alpha$ is a rotation matrix, $\mathbf{M}_\delta$ is a retardation matrix, $\mathbf{Mirr}$ is the Jones matrix describing the effect of the retro-prism and $\mathbf{F}$ describes the polarising effect of the air-glass Fresnel reflections as a function of the angle of incidence. Subscript 1 and 2 refers to the first and second liquid crystal modulator respectively.

The above Jones calculus does not take into account the attenuation caused by the grazing incidence (proportional to $\cos(\gamma_o)$, i.e. a loss of signal $\propto \cos^4(\gamma_0)$ of 75% at $\gamma_o = 45°$). Similarly it does not account for the intensity lost by the Fresnel reflections (~20%).

The next step is to generate the normalised Stokes intensities in order to plot the available SOP on the Poincaré sphere:

$$I_x = |\mathbf{J}_{out}[1]|^2,$$
$$I_y = |\mathbf{J}_{out}[2]|^2,$$
$$I_{45} = |\mathbf{J}_{45}[1]|^2,$$
$$I_{cp} = \mathbf{J}_{cp}[1]|^2$$

Where

$$\mathbf{J}_{45} = \mathbf{R}_{45°} \cdot \mathbf{J}_{out} \text{ and } \mathbf{J}_{45} = \mathbf{R}_{45°} \mathbf{M}_{\pi/4} \cdot \mathbf{J}_{out}$$

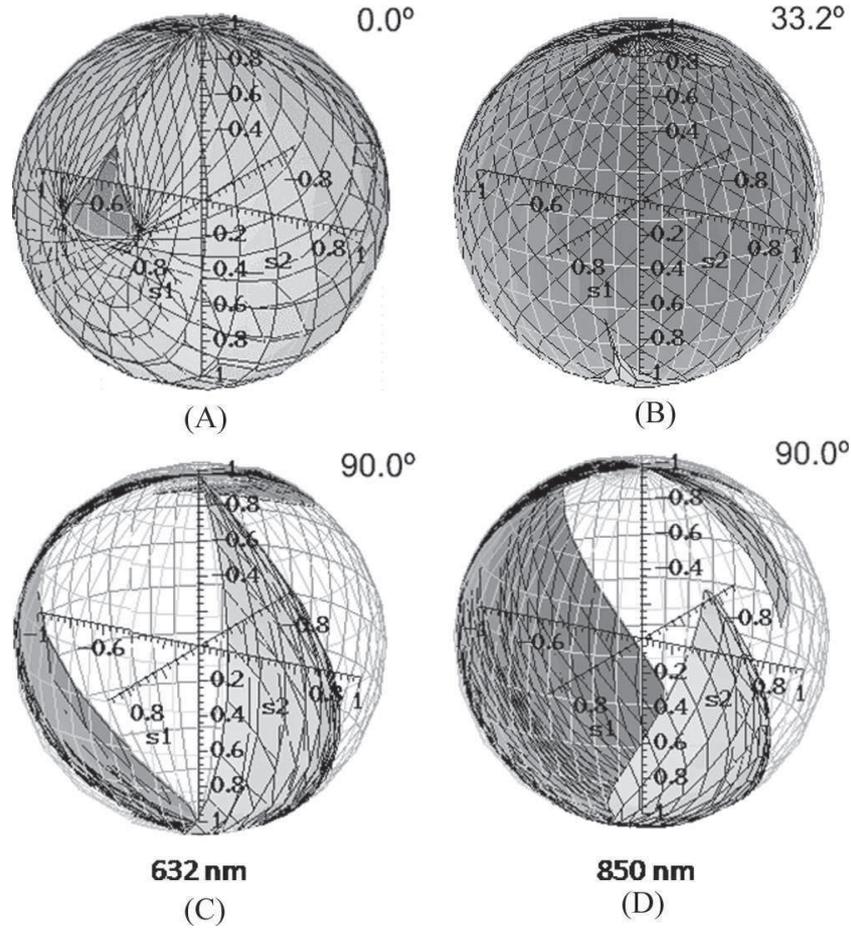

**FIGURE 5** Calculation of the variation in coverage of the Poincaré sphere as a function of the angular alignment ($\Delta\phi$) between the two cascaded liquid crystal modulators. The calculations have been made for 1.4 µm thick cells at both 632 nm ($\Delta n = 0.14$) left and 850 nm ($\Delta n = 0.10$), right. The optimum angles were determined to be 0° in the case of 632 nm (A, showing 95% coverage) and 32° for 850 nm (B, 52%), while the intuitively driven choice of 90° was shown to be inadequate for both wavelengths (C + D).

The normalised Stokes parameters are then easily calculated as:

$$S_0 = I_x + I_y,$$
$$s_1 = (I_x - I_y)/S_0,$$
$$s_2 = (S_0 - 2I_{45})/S_0,$$
$$s_3 = (S_0 - 2I_{cp})/S_0$$

The generated plots, based on the electro-optical characterisation, were used for the selection of the material employed in the final large area modulators that were subsequently mounted on the retro-prism.

The plots serves also in the optimisation of the angular offset ($\Delta\phi$) between the two liquid crystal modulators. The relative alignment between the cascaded modulators alters dramatically the Poincaré area of available SOPs (Fig. 5). The area available for modulation determines ultimately the number of distinguishable SOPs that can be implemented in the communication protocol.

It is important to notice that the choice of $\Delta\phi$ is by no means trivial. As can be seen in the figure the coverage, even at normal incidence, of the Poincaré is heavily dependent of $\Delta\phi$, which has to be chosen with care. Algorithms to optimise the choice of $\Delta\phi$ for any $\phi_0$ and $\gamma_i$ and the choice of points on the sphere to be used in the communications are the subject of a patent being filed.

## CONCLUSION

The novel concept of multi PolSK in retro-modulation using LC modulators has been demonstrated. This approach enables a simplification of the wireless optical link, by eliminating the airborne laser and tracking unit. Furthermore, eliminating the airborne laser, the power limitation of the laser disappears. Additionally the system inherently reduces the airborne payload.

The current implementation of the concept facilitates bandwidths up to 100 kbps using current V-shape materials. The concept can be extended to Gbps regime by using Pockels solid state modulators.